# Application of Time-Controlled Critical Point in Pressure Reducing Valves. A Case Study in North Spain

Andrés Ortega-Ballesteros [1], David Muñoz-Rodríguez [1], María-Jesús Aguilera-Ureña [1], Francisco Javier de los Santos-Zarco [2] and Alberto-Jesus Perea-Moreno [1,*]

[1] Departamento de Física Aplicada, Radiología y Medicina Física, Universidad de Córdoba, Campus de Rabanales, 14071 Córdoba, Spain; g02orbaa@uco.es (A.O.-B.); qe2murod@uco.es (D.M.-R.); fa1agurm@uco.es (M.-J.A.-U.)
[2] GS Inima Environment, C. Gobelas, 41, 28023 Madrid, Spain; fjsantos@inima.com
* Correspondence: g12pemoa@uco.es

**Abstract:** Potable water utilities are currently making great efforts to reduce leakage rates and assure long-term supply to the population due to the challenges of climate change, growing population and water shortage scenarios that have been on them over the last years. One of the most employed methods to reduce leakage includes the installation of pressurereducing valves along the water distribution network and the utilization of pressure management schemes. Pressure management includes different types of control models, which are applied according to the requirements of each site. The most advanced and sophisticated scheme is critical point control, which relies on a flow signal from a measuring device or online communication between the critical point and the valve. This paper proposes the utilization of a seasonal autoregressive integrated moving average, or the SARIMA model, to correlate pressure at the outlet of the valve and pressure on the critical point of the area supplied, aiming to set a fixed pressure in the critical point. The SARIMA model is developed according to historical data logged in the field and then validated. Later, the SARIMA model was tested on a real location in the village of Noja, Spain. The analysis of the field test results prove that the proposed model is feasible to be used since there is no significance difference between the target values set in the critical point and the real values measured in the field. The research proves that the SARIMA model can be used as an alternative for critical point control in water distribution networks when no flow signal is available or when communication between the critical point and the pressure reducing valve is not an option.

## 1. Introduction

The utilization of pressure reducing valves (PRV) in water distribution networks (WDN) is widely extended as a tool to reduce leakage [1–3], lower burst frequency [4], improve level of service to the customers [5] and, to a minor extent, and mainly under water scarcity situations, manage customer consumption [6,7]. Most of the water utilities across the world have implemented working plans to install this type of valves to achieve one or various of the mentioned targets [8]. Moreover, scientific production over the last years related to the employment of PRV in WDN has grown drastically [9]. Over the last years, it has also become popular the utilization of pumps as turbine (PATs) in WDN, which, in addition to the pressure reducing function, can also generate energy [10,11]. Different methods have been developed in the literature for the optimal location of PATs [12,13]. Despite more research and experience still required, PATs are identified as a clear future alternative to PRVs from both an economic and a sustainability point of view [14].

A PRV whichs a hydraulic control valvewhichch reduces upstream pressure to a stable downstream pressure regardless of flow or upstream pressure fluctuations. The literature contains several references for optimal placement and location of PRVs along the network [15,16]. The PRVs are placed in the WDN working under different schemes: fixed outlet (FO), time-controlled (TC), flow modulation (FM) and critical point control (CPC) [17], with the main objective of reducing excess pressure which generates water leakage and water losses [18].

FO is the simplest and most common scheme and only requires the installation of the PRV. It is suitable mainly for areas with uniform supply features and low headloss. The main limitation of this type of control is the inability to adapt to a variable demand pattern, common of areas with seasonal demand. The rest of the schemes, TC, FM and CPC, bring additional benefits to the operation of the network, but their installation is more complex, costly, requires additional equipment and the operator workforce must be well trained to support the installation, operation and maintenance [19].

TC consists of setting different values of pressure at the outlet of the PRV according to the time of the day. TC might be achieved by the addition of hydromechanical parts to the control trim of the PRV or by the utilization of electronic controllers. The most common application is double set-point, where pressure during the minimum consumption period, usually night time, is lowered to reduce leakage and kept at a higher value during peak demand time to satisfy customer demand. Despite this method being widely employed, the literature does not contain too many references about how to select the right settings of the valve. Ulanicki et al. [20] developed a model to calculate time schedules at the outlet of the PRV utilizing a nonlinear programming problem. The use of a genetic algorithm for this purpose was later introduced by Nicolini et al. [21].

FM pressure-controlled systems rely on a flow signal generated by a flow meter, usually a pulse output or a 4–20 mA signal [22]. This system dynamically adjusts the pressure at the outlet of the PRV according to the flow, following a set-point table introduced by the user. While FM has remarkable benefits, such as being an effective method to address background leakage [23] and the possibility to adapt to real time demand [24], it also encounters some inconveniences, mainly from an operational point of view. These restrictions relate to the unavailability of a working flow meter at the installation site: sometimes it is not an option to install a flow meter due to lack of budget or because there is not enough space at the installation site. Another inconvenience appears when the flow meter malfunctions or when there are issues for transmitting the signal, common in chambers and places prone to floodings.

CPC is the most sophisticated control scheme and aims to deliver a target pressure in the critical point or critical points of the area supplied by the PRV. The critical point can be defined as the first point in the WDN where the pressure falls below the minimum desired pressure [25]. Therefore, keeping the pressure in this point just above the minimum will secure the right level of service in the rest of the pressure-managed area. The development of this control method has run in parallel to the development of communication technologies which allow linking the PRV with the critical point of the system [26]. The control algorithms for CPC might be closed-loop [27], where there is real-time communication between the critical point and the PRV, or open-loop. The open-loop model is similar to the FM, with the difference that the model is built to deliver a fixed pressure in the critical point of the area [5].

The benefits obtained by each of the methods are discussed in the literature, defining advantages and disadvantages for each of them [28–30]. All these papers conclude that dynamic models are the most convenient under variable demand conditions. However, the cost of implementation, operation and maintenance must be assessed beforehand. Unfortunately, most of the works existing in the literature are only experimental, with just a few references of results in real WDN [31].

The aim of this work is to obtain an open-loop statistical model based on data from a time series of pressures to keep a fixed pressure at the critical point of the WDN. The SARIMA model will be used for this purpose. The main advantage of this model for pressure management (PM) is that, unlike other CPC methods, it does not need a flow meter at the installation site and it does not require the critical point of the system to be connected with the PRV. As a result, the installation and commissioning process for the utilities is easier and also lowers the investment required.

The use of statistical and mathematical techniques for the analysis of time series has been known since the 19th century. A time series is a set of observations ordered or plotted against time and conclusions drawn from them. In the early 1970s, Box and Jenkins (1976) [32] devised a statistical methodology for time series analysis, applied to different scientific branches, which were latter named as ARIMA models. The ARIMA model, in its simplest version, has only three components that give the model its name: AR for autoregressive, I for integrated and MA for moving averages. Each part is represented by a positive integer (p,d,q). AR, or autoregressive, models are used under the condition that past values condition or correlate with current values, and MA, or moving average, models take into account past residuals to correctly fit the model. The application of ARIMA models requires the time series to be stationary: mean and variance are stable over time. The SARIMA model is an extension of the ARIMA model that allows modeling non-stationary series.

The ARIMA model has been employed in numerous research fields. For example, it has been used for modeling and forecasting water consumption by Zhou et al. [33], Jain and Ormsbee [34], Wong et al. [35], Caiado [36] and Baraun et al. [37]. In the medical research field, Liu et al. [38] explored the demographic characteristics of acute hemorrhagic conjunctivitis using a SARIMA model. Saz [39] modeled and forecast the inflation rate in developing countries, focusing on Turkey and Otu et al. [40] on Nigeria. Makoni et al. [41] forecast international tourist arrivals in Zimbabwe. Divisekara et al. [42] used the SARIMA model to forecast the price of red lentils in Canada, the largest producer in the world. Martinez-Acosta et al. [43] used the model for time series from monthly rainfall and Koohfar et al. [44] for electric vehicle charging demand. Dabral and Murry [45] used SARIMA models for monthly, weekly and daily monsoon rainfall time series. Valipur [46] validated SARIMA models to forecast long-term runoff in the United States. Akipanar and Yumusak [47] applied ARIMA models to forecast gas consumption in Turkey to improve demand forecasting.

This paper describes a time-based open-loop CPC model, with SARIMA applied to pressure management (PM) in WDN. The model relates the ratio of the system critical point pressure (*P3*) to the pressure at the outlet of the valve (*P2*) as a function of time (*P3/P2*), creating a predictive model which assures target pressure in the critical point of the WDN. The model has been developed and tested on a real location in the village of Noja in northern Spain. This village is characterized by great seasonal demand variation throughout the year. Currently, Noja water utility is applying CPC based on an open-loop control model, which relates headloss between the outlet of the valve and the critical point against the flow at the inlet of the WDN. This pressure-control strategy has been identified as the most beneficial by the water utility since it allows to keep pressure at the minimum level in the WDN, while assuring the right level of service to the customers, under the challenging seasonal demand pattern of the area. However, due to the criticality of the installation, especially during summer and bank holidays when the area is full of tourists, it would be desirable to have an alternative control scheme in case of malfunction of the flow meter to maintain target pressure in the critical point regardless of the flow measurement. This fact motivates this research and the elaboration of this paper. The results show that the new time-controlled model might be applied with satisfactory results and can be a backup alternative for water utilities in situations where they cannot apply FM or traditional CPC based on flow measurements or by connecting the PRV with the *P3*.

## 2. Materials and Methods

*2.1. Description of Case Study Location and WDN*

Noja is a seaside village in the north of Spain, in the Cantabria community (Figure 1). It is located 43 km east of Santander and 75 west of Bilbao. The total population is 2650 habitants [48]. However, mainly during the summer period, bank holidays and Easter, the population grows dramatically to up to 80,000 habitants. Indeed, Noja is identified as the Spanish town with the highest percentage, 91%, of second residences over the total of residences in the village [49].

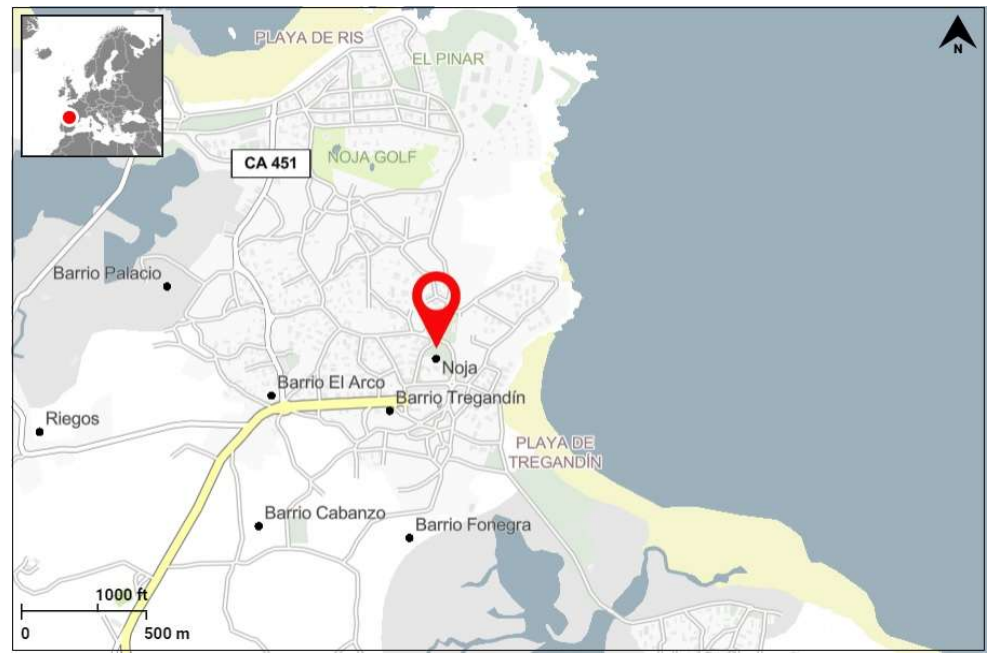

**Figure 1.** Study area location.

The population fluctuation between the summer period/bank holidays and the rest of the year generates big differences in the total amount of water supplied. Figure 2 depicts the total daily flow (TDF) delivered into the network from February 2021 to April 2022. During the months of July and August, the TDF reaches its maximum value, an average of 5080 m$^3$/day with a maximum of 6110 m$^3$/day during the Spanish long weekend in the middle of August 2021. On the other hand, the minimum TDF values occur during the winter months of January/February, with an average of 1262 m$^3$/day. During the Easter period, the TDF also increases as a result of more people traveling to the area (week of April 2022). In Figure 2, the effect of Easter is evident in 2022 but less so in 2021. This is because, in 2022, the COVID pandemic restrictions had already been lifted.

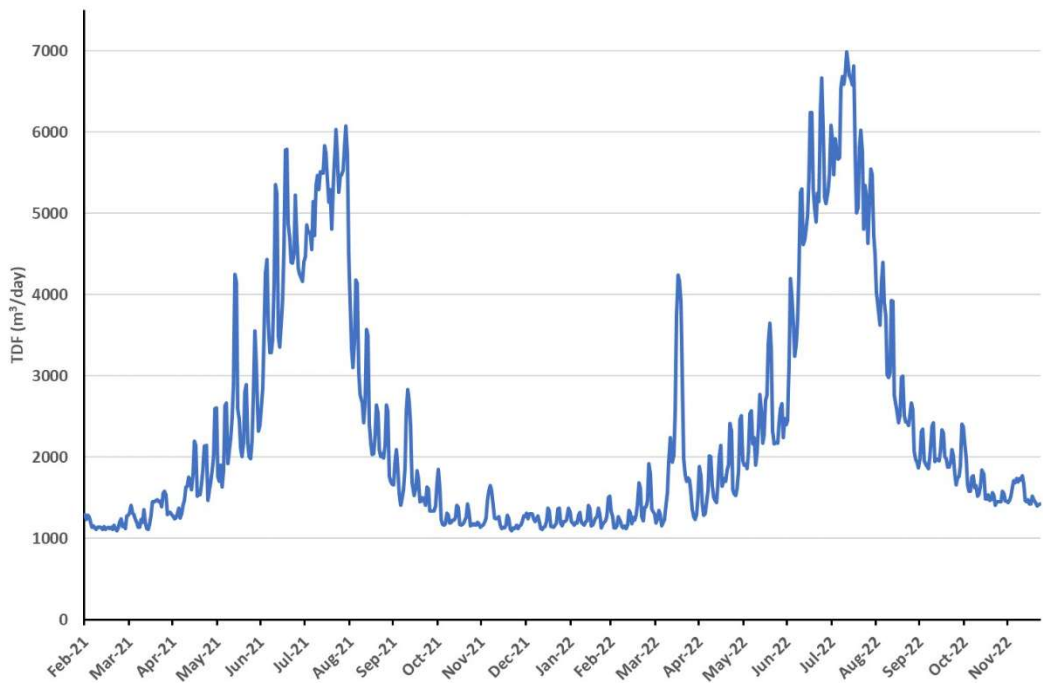

**Figure 2.** TDF from February 2021 to March 2022 in Noja village.

The water operator must deal with a very seasonal demand pattern, where the maximum demand is 539% higher than the minimum demand period. Figure 3 shows the hydraulic performance between a typical winter week of low consumption compared to a summer week of high demand. The pressure at the inlet of the PRV, *P1*, is represented by the red line. The pressure at the outlet of the PRV, *P2*, is the blue line. The pressure at the critical point, *P3*, is represented by the green line and the *flow* at the inlet of Noja is the black line.

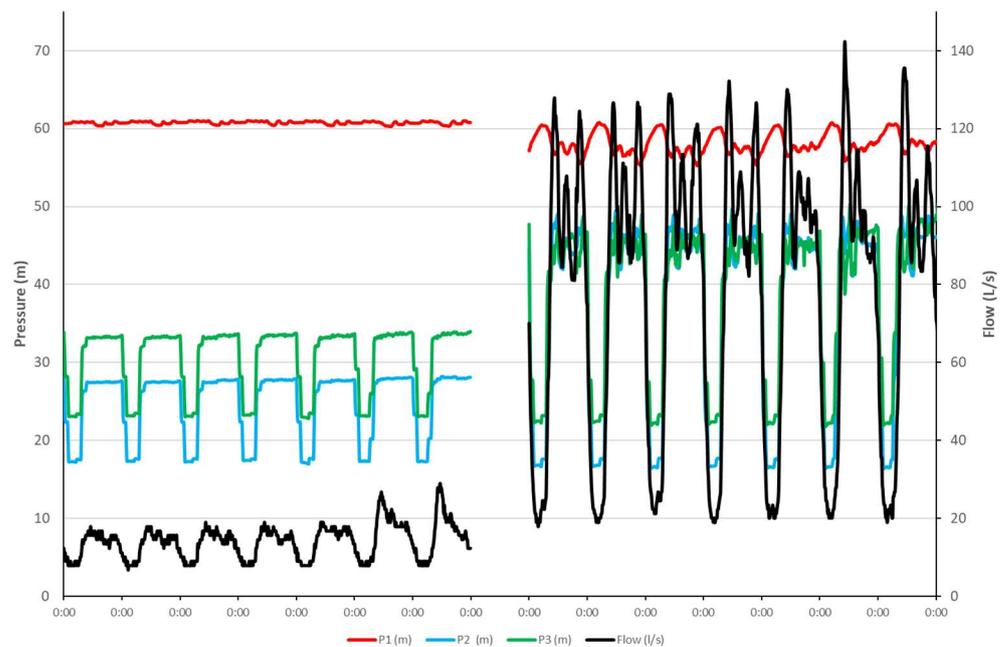

**Figure 3.** Hydraulic performance of WDN. Left: low-consumption winter week; Right: high-consumption summer week.

The situation described above with a very variable demand pattern creates a challenging scenario, which requires the employment of advanced pressure management to assure the right level of service to the customer, while keeping pressure as low as possible to reduce leakage rates and the appearance of new bursts. Since the demand is much higher during the summer period, the operator decides to keep a higher value of pressure, assuring the level of service. On other hand, during the winter, with lower demand, the pressure is much lower.

Noja town supply is fully gravitational from a distribution tank. The length of the WDN amounts to a total of 48.33 km. The main pipe materials are ductile iron (DI) and polyethylene (PE), which represent around 90% of the total length. The remaining 10% consists of asbestos (AB) and polyvinyl chloride (PVC) pipes. Large sizes, from DN150 to DN300, are made up of metal material. For sizes below DN150, plastic materials are used.

*2.2. Hydraulic Control Setup Description*

The water supply at Noja is controlled by a DN300 PRV equipped with an electronic controller which varies pressure at the outlet of the valve, *P2*, according to the flow to keep a fixed pressure at the critical point *P3*. It is an open-loop CPC method similar to the one used in [5], which relies on the flow measurement at the inlet of the WDN. The flow signal comes from the pulse output of a mechanical water meter installed upstream the PRV. The electronic controller is installed in parallel and isolated from the original pressure reducing pilot of the valve to avoid interferences over the control of the valve. The controller is logging upstream and downstream pressure at 15 min logging interval with a sampling rate of 10 s. The controller sets pressure at the outlet of the valve according to the pressure management scheme implemented. The controller works with a dead band of ±0.5 m industry standard for this type of devices. A datalogger is installed in the critical point logging pressure data with the same logging interval and sampling rate.

Below, Figure 4 shows the location of the PRV and the critical point in Noja WDN, including the elevations of each of the sites.

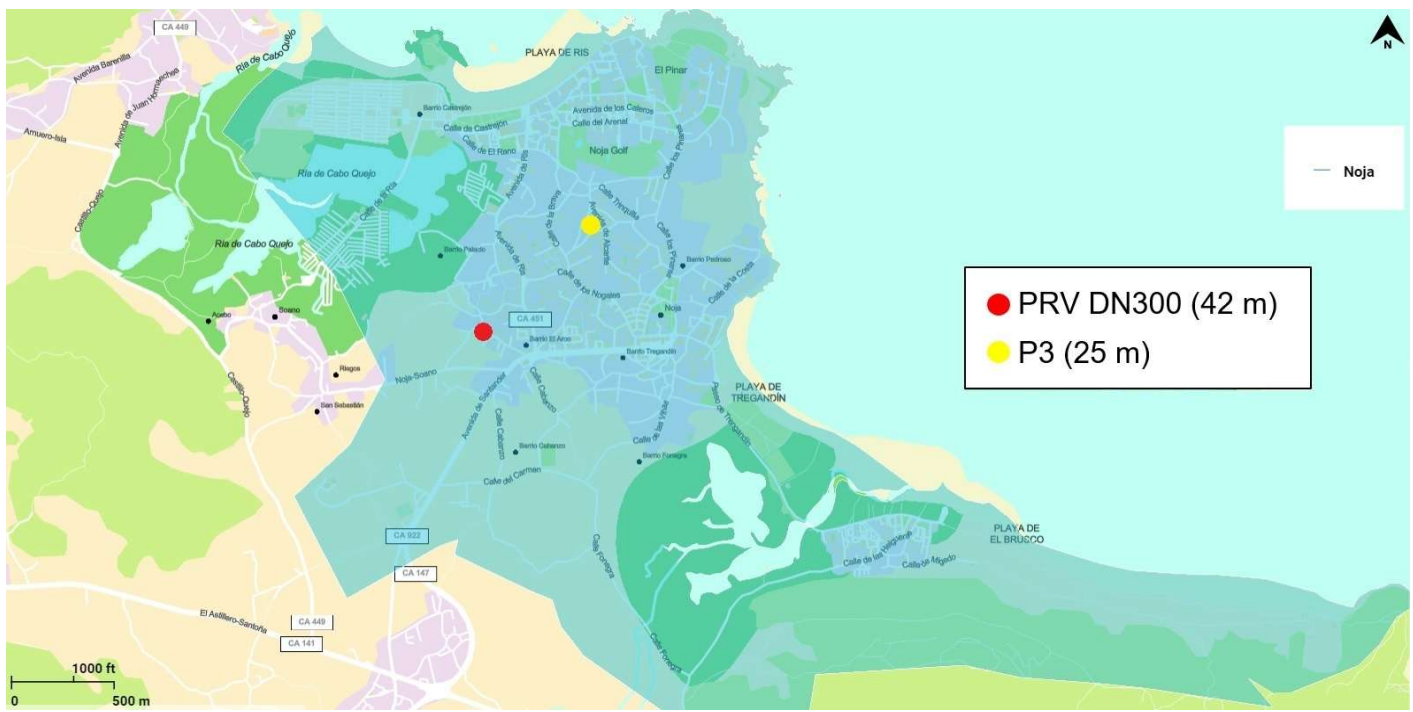

**Figure 4.** *P3* and PRV location in Noja and the area supplied by the PRV.

The critical point target pressure has been chosen by the operator according to their knowledge of the system and the area, as per Table 1 below. The operator keeps different

*P3* target values according to the time of the day, discriminating between working days and weekends.

**Table 1.** Critical point target pressure (m).

|  | Working Days (Monday to Friday) | Weekend (Saturday and Sunday) |
|---|---|---|
| 00:00 to 01:00 | 28 | 28 |
| 01:00 to 02:00 | 28 | 22 |
| 02:00 to 05:00 | 22 | 22 |
| 05:00 to 06:00 | 24 | 23 |
| 06:00 to 07:00 | 24 | 26 |
| 07:00 to 08:00 | 32 | 26 |
| 08:00 to 09:00 | 32 | 33 |
| 09:00 to 12:00 | 33 | 33 |
| 12:00 to 13:00 | 32 | 33 |
| 13:00 to 14:00 | 32 | 32 |
| 14:00 to 00:00 | 33 | 32 |

The new model proposed in this paper will target the same critical point pressure data. These values aim to reduce the pressure to the minimum possible when the demand is low to reduce leakage rates and increase during high consumption period to secure the right level of service to the customers.

### 2.3. Optimization Model

The seasonal autoregressive integrated moving average (SARIMA) model is employed specifically for data series with a seasonal component. This model is equivalent to the autoregressive integrated moving average (ARIMA) model, with the exception of the S component for the seasonal factor. The model is obtained by combining autoregressive analysis and moving averages of the values and errors due to the randomness of the time series. In this way, a value of the time series considered is related or is due to the correlation of the previous values and with the moving averages of the statistical residuals of the time series considered to obtain the model. Once the model is adjusted, forecasts are made for the modeled series of values.

Data modeling versus time is obtained following a Box–Jenkins approach consisting of four stages: model identification, parameter estimation, model diagnosis and forecast verification. The identification is performed with the autocorrelation function (*ACF*) and partial autocorrelation function (*PACF*) plots. However, the stationarity of the time series must also be checked, as it is a fundamental criterion for applying the proposed approach.

The Dickey–Fuller technique (1979) [50] is one of the most widely used methods to demonstrate the stationarity of the series. It is based on a hypothesis test where the null hypothesis indicates that the series is not stationary and has unit roots, and the alternative hypothesis indicates that the series is stationary and has no unit roots.

The autocorrelation function (*ACF*) represents the correlation of each value of the series considered with the lagged values at different lengths. The values to be estimated are those above the confidence limits. The autocorrelation is calculated using the following expression:

$$ACF(k) \; o \; AC(k) = \frac{Cov(y_t, y_{t-k})}{Var(y_t)} \tag{1}$$

where $Cov$ is a first lag correlation in the relationship of each of the values of the time series, at an instant $t$, with the values measured at the previous instant, $t-k$ and without taking into account the intermediate values when the lags are greater. This function, in ARIMA models, indicates the order (q) of the moving averages. The partial autocorrelation function (PACF) is used to find the partial correlation between the separated residuals. As in the previous function, the lag of the correlation means the relationship, in this case, of the residuals calculated at an instant t with the immediately previous ones separated at a specific time instant t-k, taking into account the intermediate residuals.

$$PACF(k) \text{ o } PAC(k) = \frac{Cov(y_t, y_{t-k}|y_{t-1}, y_{t-k+1})}{\sqrt{Var(y_t|y_{t-1}, y_{t-k+1}) \cdot Var(y_{t-k}|y_{t-1}, y_{t-k+1})}} \quad (2)$$

The selection of the best model can be performed using the Akaike criterion (AIC). This test allows selecting the SARIMA model with the best estimation and with the smallest number of parameters, so that increasing the number of parameters for a better fit is detrimental to the result of the test. The best model will be the one with the lowest value. The value is obtained by the following function:

$$AIC = -2 \ln(l) + 2k \quad (3)$$

where $k$ is the number of model parameters and $l$ is the maximum value of the likelihood function for the estimated model.

Another criterion for selecting the SARIMA model is the Bayesian information criterion (BIC).

$$BIC = k \ln(n) + n \ln(l) \quad (4)$$

where $k$ is the number of parameters to be estimated, $l$ is the error variance and $n$ is the total number of observations.

Another interesting test is the Ljung–Box test. The statistical formula of the Ljung–Box test is as follows:

$$Q = n(n+2) \sum_{j=1}^{h} \frac{\rho_j^2}{n-j} \quad (5)$$

where $n$ is the sample size, $\rho_j$ is the autocorrelation at lag $j$ and $h$ is the total number of lags tested by Ljung and Box (1978) [51].

The software packages chosen to obtain the model are IBM SPSS 28 and the EVIEWS 12 SV package. The necessity of using both softwares was the statistical demonstration of the stationarity of the data series being worked with, namely, the Augmented Dickey–Fuller test, and the use of the model verification tests, AKAIKE and *BIC*, to select the best model, as conducted by Gocheva-ilieva et al. [52] and Tabachnick and Fidell [53]. With the EVIEWS package as the stationarity test, the identification of the orders of the regular and seasonal part as well as the assessment of the different estimated models using the AIC and BIC criteria are performed. The IBM SPSS 28 software, with the expert model, is used to obtain the model coefficients, forecasts and estimators that justify the goodness of fit, as well as the Ljung–Box test.

## 3. Results and Discussion

### 3.1. Time-Based SARIMA Statistical Control Model

The available data series covers almost the entire year 2021 with hourly intervals. The period from April to September was selected to obtain a model which can be used for any date of the year. This selection also allows enough data available for model validation. Data from October to December will be used for model validation purposes.

The time series to be modeled is obtained from the quotient between the critical point pressure (*P3*) and the pressure at the outlet of the PRV (*P2*), measured every hour, between 0:00 h and 23:00 h on each of the days considered. It involves a total of 4392 datapoints.

Before identifying the model, it is analysed whether the data series is stationary or not, a condition to be fulfilled to obtain the ARIMA model. The original data series is subjected to the Augmented Dickey–Fuller test using the EVIEWS program test to confirm the stationarity or non-stationarity of the data. The result obtained has a *p*-value greater than 0.05, specifically 0.0629, which means that the null hypothesis is fulfilled; thus, the series is non-stationary and it must be transformed to make it stationary.

Figure 5 shows the original *P3/P2* series (left) and the correlogram of the original series (right). According to the representation of the correlation coefficients (AC), the presence of seasonality can be deduced by the fact that the 1st and 24th autocorrelation coefficients take maximum values, which makes regular differentiation or taking logarithms unnecessary, as the original data do not present a trend.

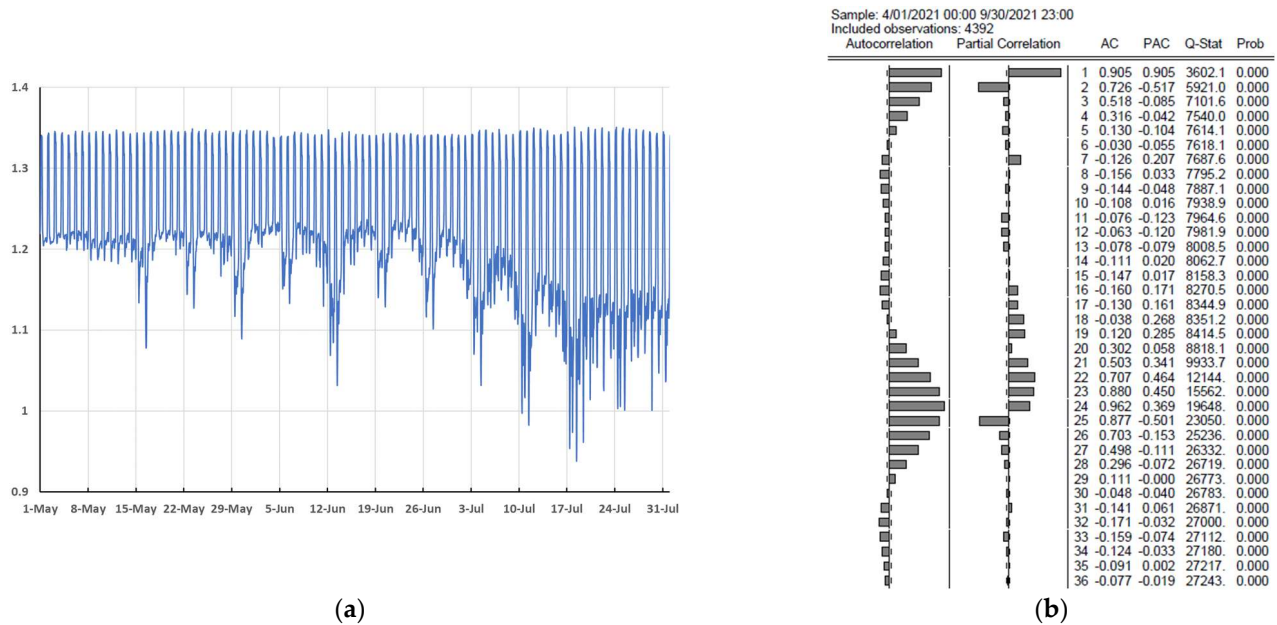

**Figure 5.** *P3/P2* original data series (**a**) and correlogram of the series without transformation (**b**).

Knowing the non-stationarity condition, and the seasonality present in the data, a seasonal difference is made to the original data series. Figure 6 represents the series obtained after differentiation by removing the seasonal component (a) and a correlogram of the series with seasonal difference (b).

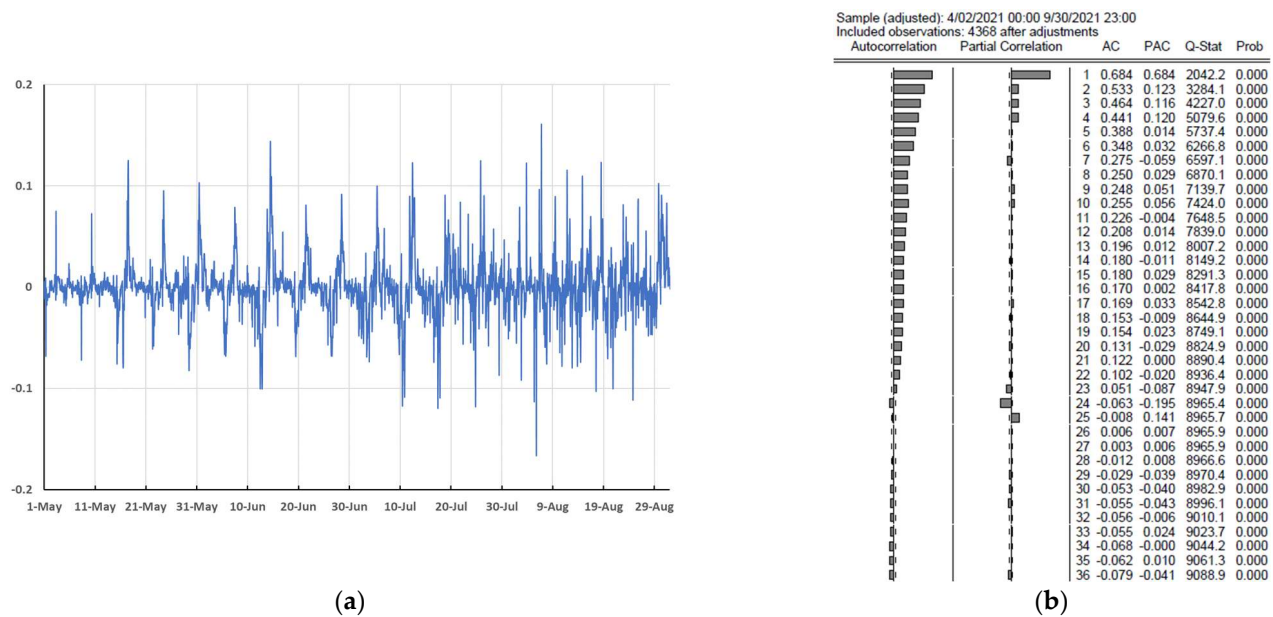

**Figure 6.** *P3/P2* series with seasonal differentiation (**a**) and correlogram of the series with seasonal difference (**b**) (EVIEWS).

The Dickey–Fuller test is then performed again, considering the cycle of 24, to check that the seasonal component has been eliminated. A significance level less than 0.05 is obtained, which leads to the conclusion that the new series is stationary, rejecting the null hypothesis. Once the stationarity has been confirmed, the model is built.

The choice of model is made using the correlograms in Figure 6 and using the criteria of AKAIKE and BIC [52,53].

From the correlogram obtained with the seasonally differenced series (Figure 6b), it is estimated that the best fitting model would be an ARIMA $(1,0,10)(1,1,1)_{24}$ pending confirmation by the aforementioned statistical criteria.

Table 2 shows a sample of the comparison of the estimated models, using the EVIEWS program, considering the information obtained from the correlogram in Figure 6b. The models shown have been compared using the differenced data series. The model and its coefficients are obtained using the IBM SPSS 28 software package based on the model estimated in the previous section [54]. The results obtained are shown in Table 3.

**Table 2.** Comparison of ARIMA models with the seasonally differentiated series.

| Model | LogL | AIC | BIC | HQ |
|---|---|---|---|---|
| (1,0,10)(1,1,1) | 12,216.502583 | −5.587226 | −5.566771 | −5.580008 |
| (1,0,9)(1,1,1) | 12,204.885563 | −5.582365 | −5.563371 | −5.575662 |
| (1,0,8)(1,1,1) | 12,203.753800 | −5.582305 | −5.564772 | −5.576118 |
| (1,0,7)(1,1,1) | 12,199.556883 | −5.580841 | −5.564769 | −5.575169 |
| (1,0,6)(1,1,1) | 12,167.589111 | −5.566662 | −5.552051 | −5.561506 |
| (1,0,5)(1,1,1) | 12,162.595805 | −5.564833 | −5.551683 | −5.560193 |
| (1,0,4)(1,1,1) | 12,157.917276 | −5.563149 | −5.551460 | −5.559024 |
| (1,0,3)(1,1,1) | 12,156.394641 | −5.562910 | −5.552682 | −5.559300 |
| (1,0,2)(1,1,1) | 12,147.705222 | −5.559389 | −5.550622 | −5.556295 |
| (0,0,10)(1,1,1) | 12,137.906410 | −5.551697 | −5.532703 | −5.544994 |
| (0,0,8)(1,1,1) | 12,128.534052 | −5.548321 | −5.532249 | −5.542650 |
| (0,0,9)(1,1,1) | 12,128.534099 | −5.547864 | −5.530330 | −5.541676 |
| (0,0,7)(1,1,1) | 12,122.037405 | −5.545805 | −5.531194 | −5.540649 |
| (1,0,1)(1,1,1) | 12,113.475700 | −5.544174 | −5.536868 | −5.541596 |
| (0,0,6)(1,1,1) | 12,086.507852 | −5.529994 | −5.516845 | −5.525354 |
| (1,0,0)(1,1,1) | 12,075.717856 | −5.527343 | −5.521499 | −5.525281 |
| (1,0,10)(0,1,1) | 12,058.167525 | −5.515187 | −5.496192 | −5.508484 |
| (1,0,9)(0,1,1) | 12,046.081398 | −5.510111 | −5.492577 | −5.503923 |
| (1,0,8)(0,1,1) | 12,043.369255 | −5.509327 | −5.493255 | −5.503655 |
| (1,0,7)(0,1,1) | 12,038.038847 | −5.507344 | −5.492733 | −5.502188 |
| (1,0,6)(0,1,1) | 12,003.081882 | −5.491796 | −5.478646 | −5.487155 |
| (1,0,5)(0,1,1) | 11,992.964592 | −5.487621 | −5.475932 | −5.483496 |
| (1,0,4)(0,1,1) | 11,987.802760 | −5.485716 | −5.475488 | −5.482106 |
| (1,0,3)(0,1,1) | 11,986.369370 | −5.485517 | −5.476751 | −5.482423 |
| (0,0,5)(1,1,1) | 11,983.222557 | −5.483161 | −5.471472 | −5.479036 |
| (1,0,2)(0,1,1) | 11,976.654733 | −5.481527 | −5.474221 | −5.478949 |
| (0,0,10)(0,1,1) | 11,966.175947 | −5.473524 | −5.455991 | −5.467336 |
| (0,0,8)(0,1,1) | 11,954.957216 | −5.469303 | −5.454692 | −5.464147 |

Table 3. SARIMA model coefficients.

| VARIABLE | Coefficient | STD.ERROR | t-Statistic | Prob |
|---|---|---|---|---|
| AR (1) | 0.974 | 0.006 | 152.297 | 0.000 |
| SAR(24) | 0.309 | 0.019 | 16.327 | <0.001 |
| MA (1) | 0.328 | 0.016 | 20.047 | <0.001 |
| MA (2) | 0.183 | 0.016 | 11.221 | <0.001 |
| MA (3) | 0.08 | 0.016 | 4.982 | <0.001 |
| MA (5) | 0.032 | 0.016 | 2.043 | 0.041 |
| MA (7) | 0.123 | 0.016 | 7.768 | <0.001 |
| MA (8) | 0.069 | 0.016 | 4.358 | <0.001 |
| MA (10) | −0.074 | 0.015 | −5.009 | <0.001 |
| SMA(24) | 0.873 | 0.019 | 87.555 | 0.000 |

An $R^2$ value of 0.973 indicates the goodness of fit and a value of 12.098 of the Ljung–Box Q with a significance level of 0.147 confirms the randomness of the residual errors.

Figure 7 shows the plot of the model fit with the data considered in the modeling. To facilitate the visibility of the data, only two months have been represented: April, a month of low consumption (a), and August, a month of high consumption (b). It can be seen that the values obtained by the model for the *P3/P2* ratio fit very well to the real data.

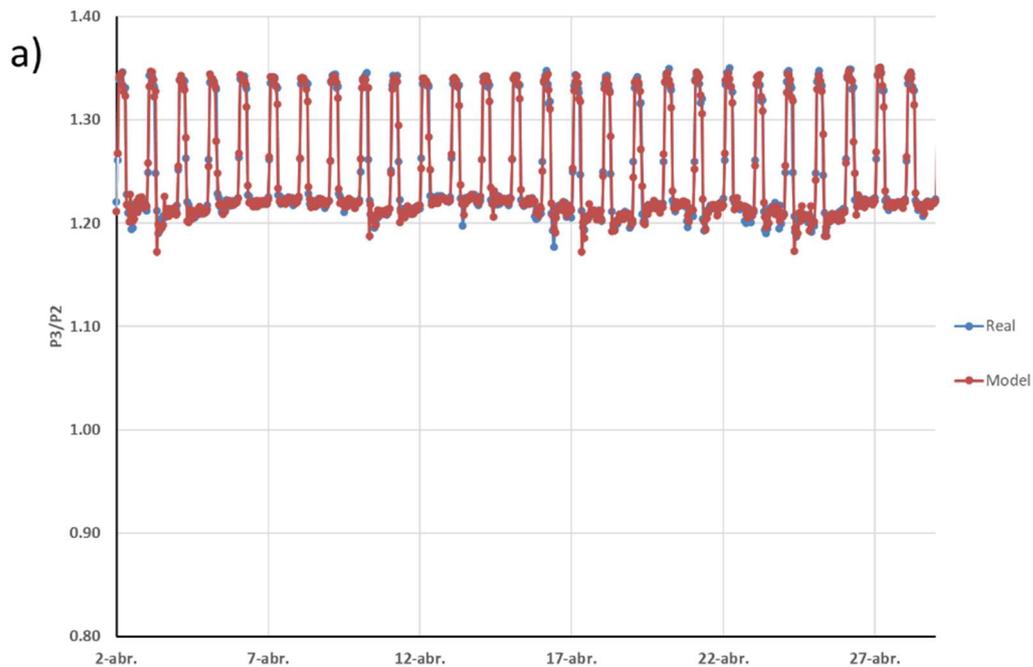

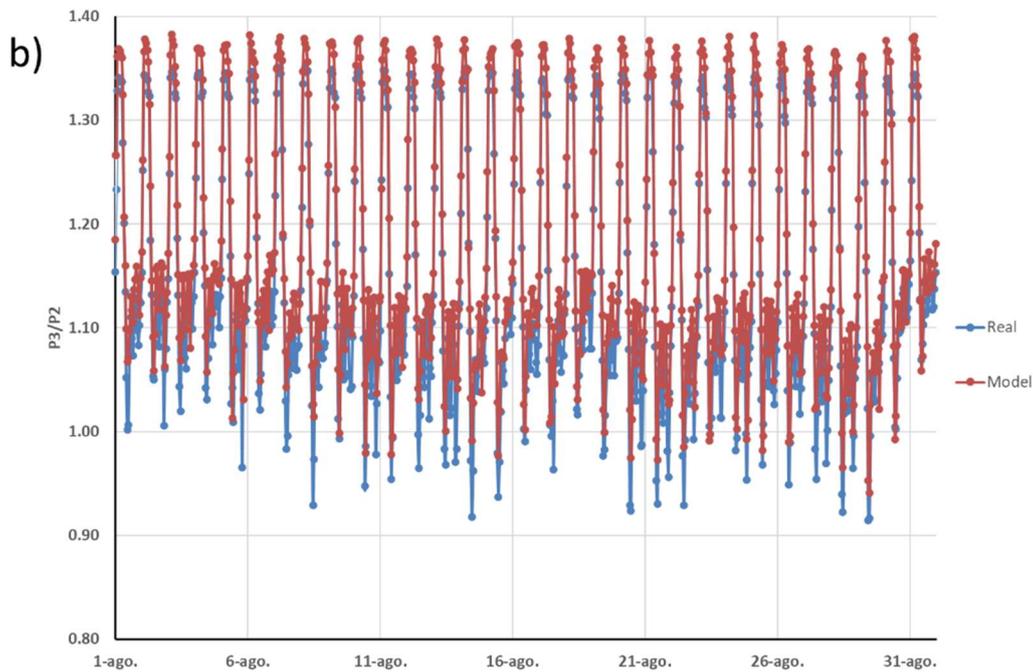

**Figure 7.** Representation of model-derived values and real values of *P3*/*P2*: (**a**) month of April, (**b**) month of August.

Once the model has been obtained, it is validated by applying it to data which were not used to build the model, from October to December 2021. Figure 8 shows the modeled and actual *P3*/*P2* values. For ease of visualization, only the month of October is shown. The RMSE is also calculated between the real values measured at the Noja facility between 1 October 2021 and 31 December 2021 and the predictions made for the same dates, obtaining a value of 0.018.

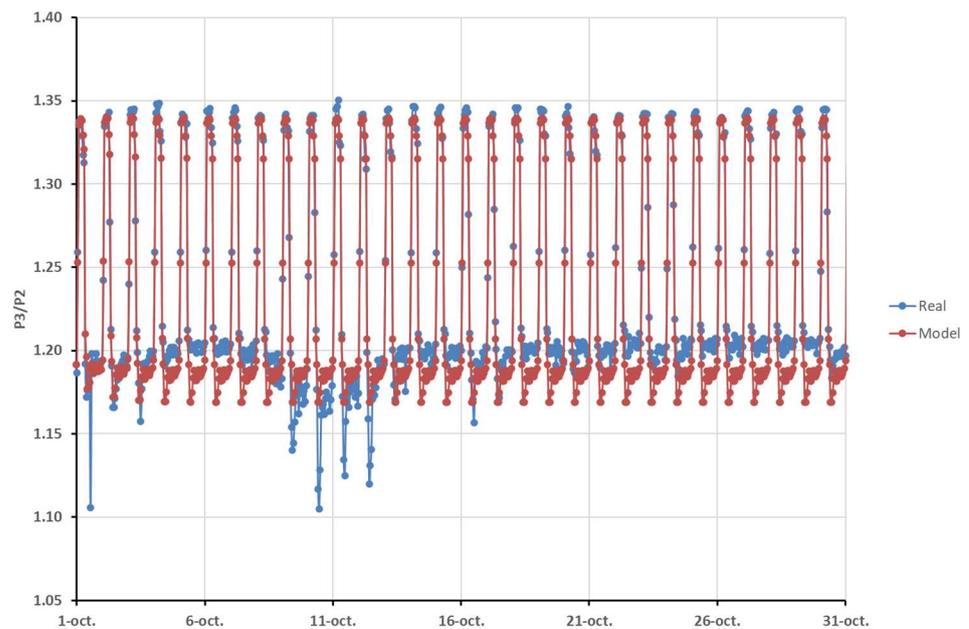

**Figure 8.** Representation of actual and derived values of the *P3*/*P2* model in October (Validation).

Figure 9 shows the fit of the prediction with the values sampled at the installation from 1 October to 31 December. The value of the coefficient of determination, 0.9648, confirms the good correlation between the actual values and the predictions obtained with the model [55].

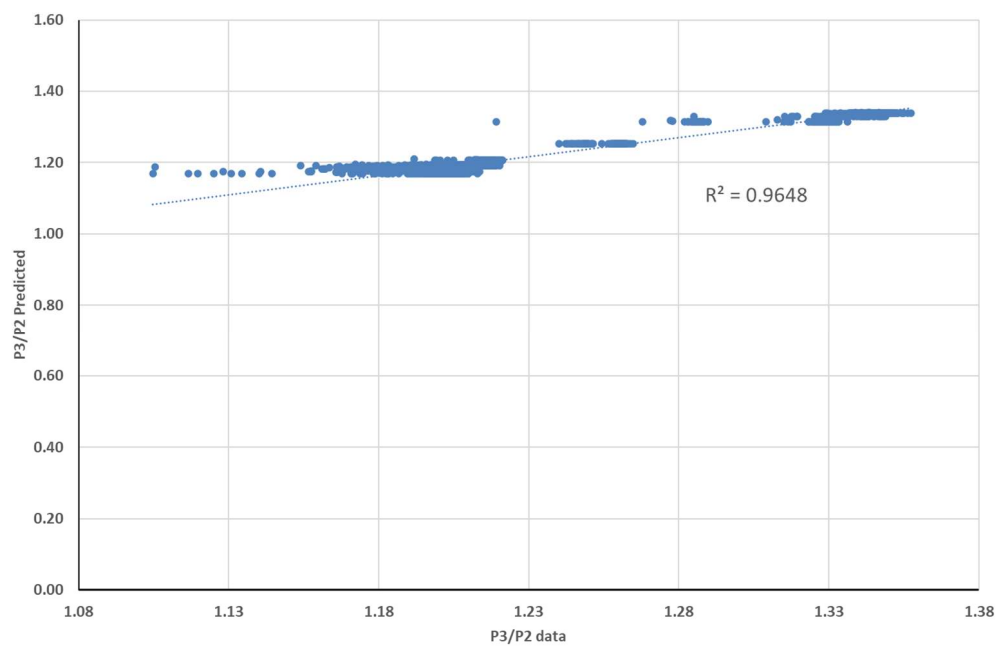

**Figure 9.** Coefficient of determination between actual and predicted values.

The new SARIMA model allows to predict the headloss between the PRV and the critical point of the area at any give day/time, based on previous historic pressure data of the area. As an advantage to the model built in [5], flow measurement is not needed. Unlike the model built by Fontana et al. [27], the SARIMA model will permit commissioning a CPC PM strategy without connecting the PRV with the critical point of the WDN.

*3.2. Hydraulic Data Analysis*

The new time-based pressure management SARIMA model to keep a target pressure in the critical point of the area has been applied on real operation for 12 days, from 12 December to 23 December 2022. Starting from critical point pressure values defined by the water utility operator, the *P3* target, shown in Table 1, and applying the model, the pressures *P2* at the outlet of the PRV were calculated and adjusted. By setting these pressures *P2* on the controller commanding the PRV, the pressures at the critical point *P3* were measured. Figure 10 describes the performance of the WDN under the SARIMA model, compared to the former pressure management scheme for CPC based on the flow measurement. In order to make the comparison, a similar period of days, between 21 November and 2 December, just before the SARIMA model was implemented, was chosen.

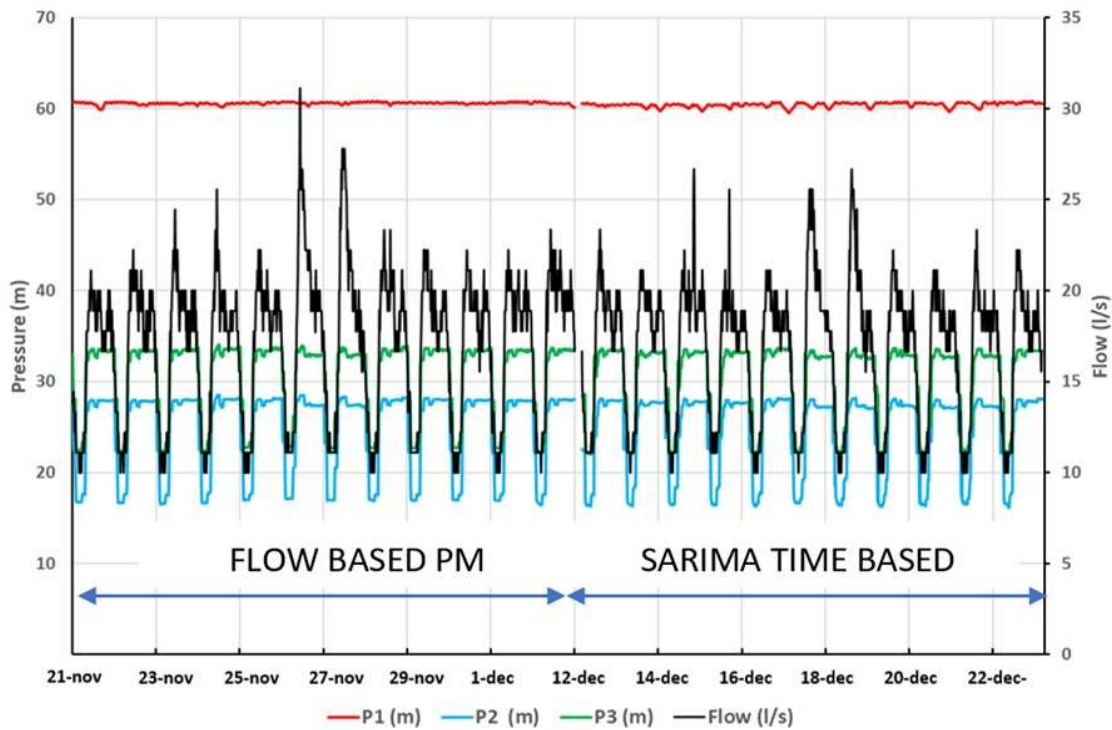

**Figure 10.** WDN performance under critical point time-based pressure management model.

Under the new SARIMA working scheme, the pressure at the outlet of the PRV, *P2*, changes according to the day and the time of the day to keep the defined target pressure in the critical point, *P3*. Under the previous flow-based PM scheme, *P2* varies according to the instantaneous flow measurement following a control model which relates headloss between *P2* and *P3* and the *flow* with the same target to keep a fixed pressure value in the critical point. The flow signal is not needed when applying the SARIMA PM model and it is shown in the graph just for illustration purposes.

The average, maximum and minimum pressure *P1*, *P2*, *P3* and *flow* during the two phases described in Figure 10 are summarized in Table 4. The data in SARIMA PM show between brackets the difference compared to the previous phase in absolute value and in percentage. The table shows that there is almost no difference in the hydraulic variables of the area when the new SARIMA PM model is applied.

Table 4. Hydraulic data under SARIMA and flow-based PM schemes.

|  | FLOW-BASED PM | | | SARIMA PM | | |
| --- | --- | --- | --- | --- | --- | --- |
|  | Maximum | Minimum | Average | Maximum | Minimum | Average |
| *P1* (m) | 60.84 | 59.85 | 60.59 | 60.82 (0.02; 0.03%) | 59.51 (0.34; 0.57%) | 60.42 (0.17; 0.28%) |
| *P2* (m) | 28.54 | 16.45 | 24.96 | 28.19 (0.35; 1.24%) | 16.15 (0.30; 1.86%) | 24.67 (0.29; 1.18%) |
| *Flow* (l/s) | 31.11 | 10.00 | 17.09 | 26.67 (4.44; 16.65%) | 9.99 (0.01; 0.10%) | 16.75 (0.34; 2.00%) |
| *P3* (m) | 34.07 | 22.18 | 30.55 | 33.80 (0.27; 0.80%) | 21.89 (0.29; 1.32%) | 30.27 (0.28; 0.93%) |

*3.3. P3 Target Pressure Evaluation*

The aim of the new model proposed is to be used as an alternative to the current open-loop critical point control based on the flow signal at the inlet of the WDN. Figure 11 shows the difference between the target *P3* pressure and the real measurement obtained after implementing the new model.

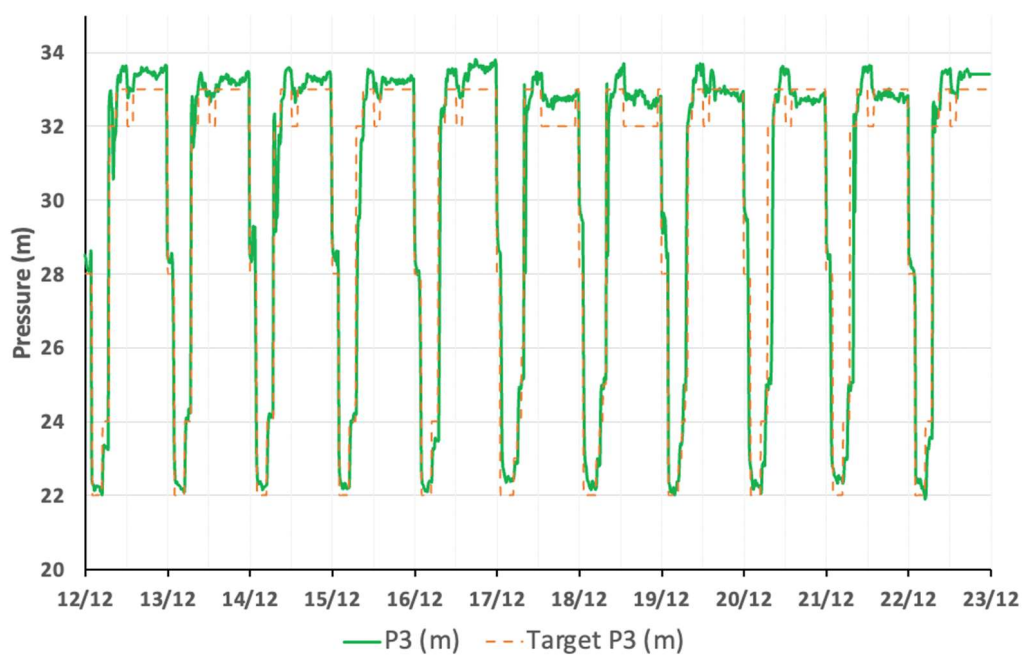

**Figure 11.** *P3* target versus *P3* real measurement.

Since data are logged every 15 min for 11 days, there are a total of 1056 datapoints during the whole measurement period. The measured *P3* value is higher than the target in 770 datapoints, 73% of the total, and below the target in 286 datapoints, 27% of the total.

When *P3* is above the *P3* target, the average deviation is 0.53 m. When *P3* is below the *P3* target, the average deviation is −0.85 m. A total of 89% of the measurements fall in the range of ±1 m deviation from the target, which is considered acceptable for the purpose of this model. Figure 12 shows the histogram with the deviation per range.

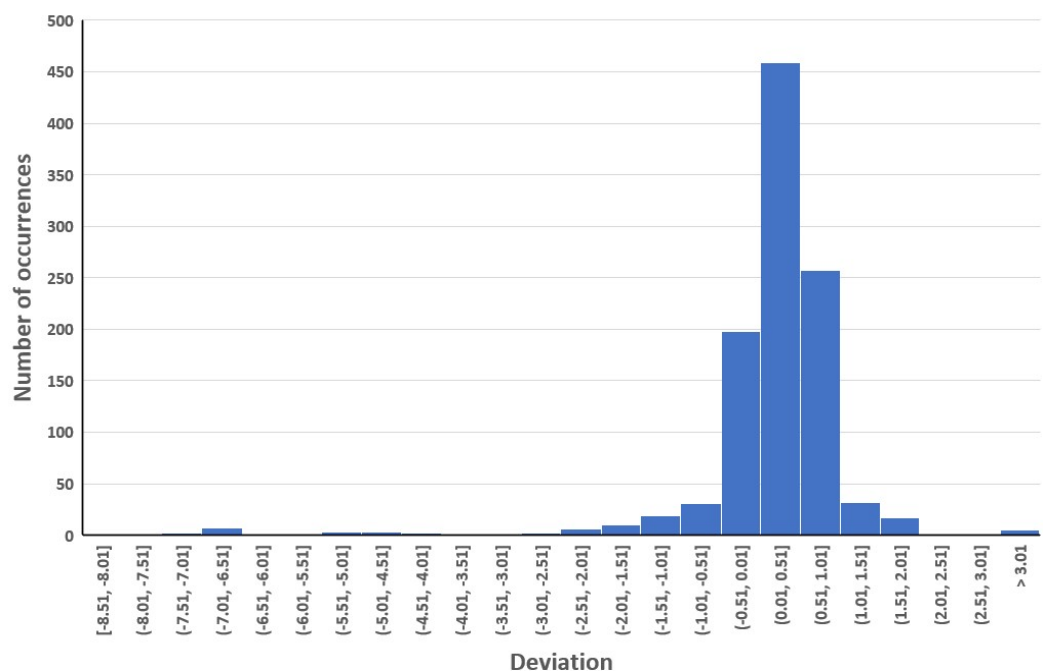

**Figure 12.** Histogram with deviation from the target pressure.

Figure 13 represents the value of target and measured *P3* pressures by box plots. To confirm what the box plot seems to indicate, an ANOVA test is performed, comparing both sets of pressures. A *p*-value of 0.412 is obtained, indicating that there is no significant difference between the two sets of pressures.

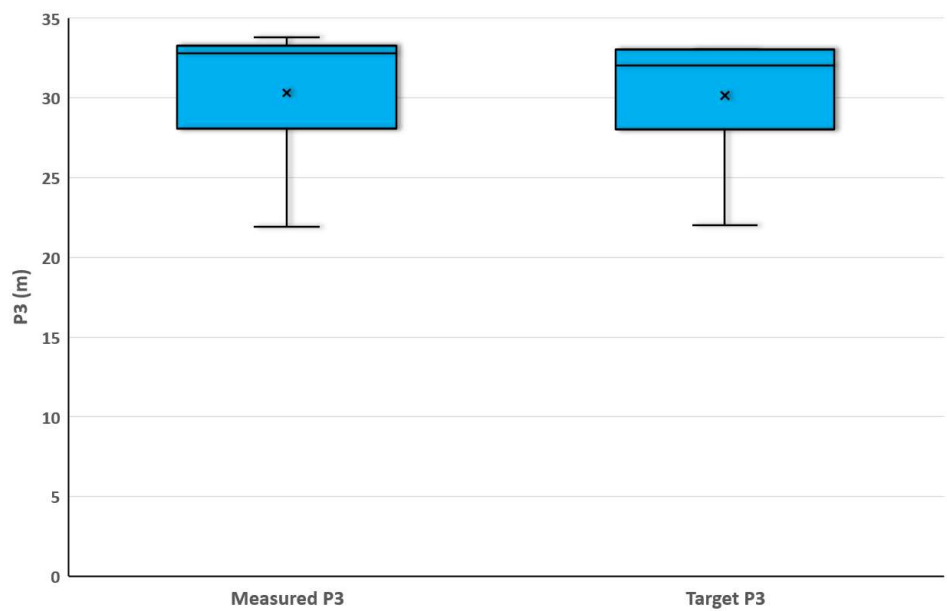

**Figure 13.** Box plot of real measured values and target *P3* values.

## 4. Conclusions

PM includes a different control model for the optimization of WDN. CPC is identified as the most advantegeous method under variable and seasonal demand pattern since it allows to reduce leakage in the network while providing the right level of service to the customer. However, current CPC methods rely on existing flow measurements or on the connection of the PRV with the critical point of the system. Sometimes, these two issues are not possible from a technical point of view or imply massive cost.

In this work, a time series of a ratio of pressures measured downstream of the PRV and the critical point on a real location in Noja village was modeled to create and validate a new open-loop control scheme for CPC based on time. Hourly pressure values were taken from February to December 2021. Due to the influence in the demand of the holiday periods in the area, it is considered to estimate, for the modeling, data from three months of a non-holiday season, with lower consumption (April, May, June), and another three months of a holiday period, with higher consumption (July, August, September), in order to obtain a useful model regardless of the time of the year. In this way, 4392 observations were considered to estimate the model. The data from 1 October to 31 December, precisely 2208 observations, were used to validate the model.

A SARIMA model has been built and implemented as an alternative to the current PM model based on the flow reading that the Noja utility is using. The SARIMA model has been validated by comparing the values predicted by the model with the data taken between October and December 2021. The first test was conducted by calculating the RMSE, giving a value of 0.018. The coefficient of determination of the comparison was also calculated, giving a value of 0.9648. Both tests support the goodness of fit of the model to predict pressure values at the valve outlet and at the critical point of the zone. The SARIMA model was also tested in real field conditions for eleven days in December 2022. The hydraulic variables of the area were analyzed, showing that there is no difference when working under the previous model and that the PRV is under the SARIMA model control. The real field measurements at the critical point obtained when applying the SARIMA model were compared with the critical point target pressure by means of an ANOVA analysis with a *p* value of 0.412, greater than 0.05, thus confirming the null hypothesis of this analysis and that there is no statistical difference between the values compared.

The result of the field test and the statistical analysis performed concludes that SARIMA is a useful tool to modulate pressure at a PRV outlet to achieve a target value at the critical point of the area supplied. The new model can be implemented as an alternative for CPC in situations where no flow signal is available or where it is not possible to establish real-time communication between the critical point and the PRV.

**Author Contributions:** Conceptualization, A.O.-B., D.M.-R., M.-J.A.-U.; F.J.d.l.S.-Z. and A.-J.P.-M.; methodology, A.O.-B., D.M.-R., M.-J.A.-U.; F.J.d.l.S.-Z. and A.-J.P.-M.; investigation, A.O.-B., D.M.-R., M.-J.A.-U.; F.J.d.l.S.-Z. and A.-J.P.-M.; resources, A.O.-B., D.M.-R., M.-J.A.-U.; F.J.d.l.S.-Z. and A.-J.P.-M.; writing—original draft preparation, A.O.-B., D.M.-R., M.-J.A.-U.; F.J.d.l.S.-Z. and A.-J.P.-M. All authors have read and agreed to the published version of the manuscript.

**Funding:** This research received no external funding.

**Institutional Review Board Statement:**

**Informed Consent Statement:**

**Data Availability Statement:**

**Acknowledgments:** The authors would like to thank GS Inima water utility for their contribution, availability, criticism, and support to materialize this paper.

**Conflicts of Interest:** The authors declare no conflict of interest.

**Abbreviations**

The following abbreviations have been used in this document:

| | |
|---|---|
| PRV | Pressure reducing valve |
| WDN | Water distribution network |
| PATs | Pumps as turbines |
| FO | Fixed outlet |
| TC | Time-controlled |
| FM | Flow modulation |
| CPC | Critical point control |
| PM | Pressure management |
| TDF | Total daily flow |